

\font\bigbrm=cmbx12



\def\simls{\hbox{$\,$\raise.5ex\hbox{$<$}
         \kern-1.1em \lower.5ex\hbox{$\sim$}$\,$}}
\def\simgt{\hbox{$\,$\raise.5ex\hbox{$>$}
         \kern-1.1em \lower.5ex\hbox{$\sim$}$\,$}}
\def\half{{\textstyle{1\over2}}}

\def\r#1{[#1]}
\def\Smoot{1}
\def\Davis{2}
\def\Bra{3}
\def\Olive{4}
\def\Copeland{5}
\def\Adams{6}
\def\AbbWis{7}
\def\Linde{8}
\def\Coleman{9}
\def\Numerical{10}
\def\Bevington{11}
\def\Hodges{12}

 \magnification = \magstep1
  \hsize=6.5truein
  \vsize=9.0truein


{\nopagenumbers
 \parindent=0pt
 \leftskip=20pt

\vfil
{\bigbrm\centerline{Constraints on Chaotic Inflation
from the {\it COBE DMR} Results}}

\vfil
\centerline{{\bf Hannu Kurki-Suonio}
 \footnote{${}^1$}{hkurkisu@pcu.helsinki.fi}
}
\smallskip
{\it\centerline{Department of Theoretical Physics}
    \centerline{University of Helsinki}
    \centerline{00014 Helsinki, Finland}
}
\bigskip
\centerline{and}
\bigskip
\centerline{{\bf Grant J. Mathews}
 \footnote{${}^2$}{mathews@physics.llnl.gov}
}
\smallskip
{\it\centerline{University of California}
    \centerline{Lawrence Livermore National Laboratory}
    \centerline{Livermore, California 94550}
}
\vfil
\centerline{\bf Abstract}
\medskip
\parindent=20pt

We explore constraints on various forms for the effective potential
during inflation based upon a statistical comparison between
inflation-generated
fluctuations in the cosmic microwave background temperature and the {\it COBE
DMR}
results. Fits to the {\it COBE} $53A+B \times 90 A+B$ angular correlation
function
are obtained using simple analytic forms for the effective potential in chaotic
inflation
models.  Not surprisingly, these fits are optimized for a nearly scale-free
fluctuation power spectrum.
However, from the $\chi^2$ distribution for the fits we can set upper limits of
$n \le 1.2$ and $n \le 7$ at the $1 \sigma$ and $2 \sigma$ confidence levels,
respectively,
for a $V(\phi) = \lambda \phi^n$ effective potential.
  Similarly, new limits on parameters for polynomial effective potentials
can be determined at the $1 \sigma$ and $2 \sigma$ confidence levels.
The most stringent constraint, however, is on
the overall magnitude of the effective potential.

\vskip .2 in
\noindent {\it PACS number(s): 98.80.Cq, 98.80.Dr, 98.80.Es}\hfil\break

\vfil\eject
}


\bigskip
\centerline{\bf I. INTRODUCTION}
\medskip

Measurements of the large-scale anisotropy in the cosmic microwave background
\break
(CMB)
by the Cosmic Background Explorer ({\it COBE}) Differential Microwave
Radiometer ({\it DMR})
experiment have provided important support for the hot big bang model
with inflation (e.g. \r\Smoot - \r\Davis).  A favored explanation for the
generation of
the observed fluctuations in the CMB temperature is by the expansion
of quantum fluctuations of a scalar field during the
inflationary epoch \r\Bra.  The fact that the observed angular correlation
function
is more or less consistent with a scale-free Harrison-Zel'dovich spectrum of
power on various angular scales is consistent with the predictions of
inflationary
models.  This is particularly true since the fluctuations resolved by
{\it COBE} are larger than the horizon at
recombination and not yet distorted from the inflation-generated
spectrum by gravitational clustering on subhorizon scales.

However, the {\it COBE} observations are not exactly scale free (i.e.
preferring
a power law index which deviates slightly from unity), and
neither are the predictions of inflation.  This is because inflation occurs
as the universe is rolling down the effective potential. Thus,
the amount of inflation is slightly different for
different angular scales. The change in
angular anisotropy of the cosmic microwave background
from scales of a few degrees to the full sky is, therefore,
a measurement of the rate of change in the effective potential during that
short
interval of inflation for which those angular scales were stretched beyond the
apparent horizon.  That the amplitude of the fluctuations depends upon the
overall amplitude of the effective potential is well known \r\Olive.  In this
paper
we discuss how the functional form of the potential is further constrained from
the observed angular correlation function.

Indeed, it is in principle possible to even reconstruct the inflation
effective potential from a knowledge of the fluctuation spectrum \r\Copeland.
However,
such a reconstruction is subject to large uncertainties and is probably
not possible in the foreseeable future \r\Copeland.  In this paper, therefore
we take a different approach.  That is, we consider specific analytic forms for
the
potential and quantify the range of parameters for these potentials which are
statistically compatible with the observations.  In this way,
we can use the {\it COBE} angular correlation
function to make a quantitative
assessment of  possible inflation-generating potentials.
Our work also differs from previous studies (e.g. \r\Copeland, \r\Adams) in
that we are using the
{\it COBE} data only.
That is, we do not consider
the large scale clustering of galaxies which also contains information on the
angular
correlation on yet smaller scales.  However, since fluctuations on smaller
scales have
experienced gravitational clustering, this generalization requires a knowledge
of the influence of dark-matter components.  By restricting ourselves to the
{\it COBE} data alone, we avoid the necessity to make
any assumptions about the nature of dark matter.  We also avoid
any possible confusion from the mixing of dissimilar data sets.

Moreover, the shape of the
inflation effective potential is not known.  A large
number of different models, with different potentials,
some with many free parameters, have been proposed, e.g. \r\Olive.
While most models lead to a spectrum which is nearly
scale invariant, this scale-invariance is not exact,
and some models lead to significant deviations from scale invariance.
Of course, the constraint which can
be placed upon effective potentials is limited
by the uncertainties in the observed correlation function, and
effects of cosmic variance \r\AbbWis. Nevertheless,
these data represent the only
direct observation of the shape of the inflation effective potential, and
the statistical analysis of these data as described here provides
at least some information as to which inflation models can be excluded.

For this study we consider chaotic inflation \r\Linde~ in particular,
with two simple forms for the effective potential,
$$\eqalignno{
   V(\phi) &= {1\over n!}\lambda\phi^n &(1)\cr
\noalign{\hbox{and}}
   V(\phi) &= \lambda\bigl({\textstyle {1\over8}\beta\phi^2
      + {1\over3}\alpha\phi^3 + {1\over4}\phi^4}\bigr), &(2)\cr}
$$
where we use Planck units, $c = m_{Pl} = 1$.  As a function of $n$ (or
$\alpha,\beta$) and $\lambda$, we do a search to find the minimum $\chi^2$ and
projected confidence limits for the parameters,
i.e.~ we use the goodness of fit to the {\it COBE} correlation function
as a criterion to constrain the degree to which the parameters of these models
can
be excluded.

\bigbreak
\centerline{\bf II. METHOD}
\medskip

The generation of the CMB anisotropy in inflationary models has been discussed
extensively in the literature (e.g. \r\Bra, \r\Olive).  A brief outline of how
the results
in this paper were calculated is as follows.

The equations governing the evolution of the scalar field and the universal
expansion are,
$$
   \ddot\phi + 3H\dot\phi + V'(\phi) = 0 \eqno(3)
$$
and
$$
   H^2 \equiv \biggl( {\dot R \over R} \biggr)^2 = {8\pi\over3}\bigl[
      V(\phi) + \half\dot\phi^2 \bigr]. \eqno(4)
$$
For simplicity, we assume instantaneous reheating as the scalar field
approaches
the minimum of the potential.

The perturbations responsible for the CMB anisotropy originate from quantum
fluctuations which are stretched beyond the apparent horizon during inflation.
Two kinds of fluctuation can contribute to the CMB.
Scalar fluctuations (i.e. fluctuations of the inflaton $\phi$) become
density perturbations.  Tensor fluctuations (i.e. fluctuations of spacetime
curvature) become gravitational waves.  The waves of interest here have
wavelengths of order the horizon scale.  The amplitudes for a multipole
expansion
of the CMB anisotropy can be written,
$$
   \langle a_l^2\rangle = \langle a_l^2\rangle_S + \langle a_l^2\rangle_T \ \ ,
   \eqno(5)
$$
where the scalar contribution is
$$
   \langle a_l^2\rangle_S = {2l+1\over25\pi}\int_0^{\omega_{\rm max}}
      {d\omega \over\omega} j_l(\omega)^2 {H^4\over\dot\phi^2}, \eqno(6)
$$
and the tensor contribution is
$$
   \langle a_l^2\rangle_T = 36l(l-1)(l+1)(l+2)(2l+1)\int_0^{\omega_{\rm max}}
      d\omega \omega F_l(\omega)^2 H^2, \eqno (7)
$$
where from ref. \r\AbbWis
$$
   F_l(\omega) = \int_0^{s_{\rm dec}} ds \biggl\lbrace -{j_2\bigl[\omega(1-s)
      \bigr]\over\omega(1-s)} \biggr\rbrace \biggl[
      {2\over(2l-1)(2l+3)}j_l(\omega s)$$
      $$+ {1\over(2l-1)(2l+1)}
      j_{l-2}(\omega s) + {1\over(2l+1)(2l+3)}j_{l+2}(\omega s)\biggr].
   \eqno(8)
$$
where
$\omega_{max} = 2(1 + Z_{eq})^{1/2}$ and
$s_{dec} = 1 - (1 + Z_{dec})^{-1/2}$.
The integrals in eqs. (5-7) are over scales $\omega \equiv kr_0$, where $k$ is
the
comoving wavelength of the perturbation, and $r_0 = 2/H_0$ is the radius of
the presently observable universe.
We used $H_0 = 50$ km/s/Mpc and the values $Z_{\rm eq} = 10500$ and
$Z_{\rm dec} = 1100$ for the redshifts of matter-radiation equality and
of hydrogen recombination, respectively.   The quantities $H^4/\dot\phi^2$ and
$H^2$ are evaluated at the epoch during inflation, when the scale in question
exits the
horizon $(k = H)$.

It should be noted, that the use of $H^2/\dot\phi^2$ and $H^2$ in the
integrands for the fluctuation amplitudes is based upon perturbation theory
results obtained for exponentially expanding spacetimes, and is thus an
approximation.  However, the solution of Eqs.~(3) and (4), as well as the
integrals
(6), (7), and (8), were done numerically.  Thus, we do not make use of the
`slow roll over' approximation.

To compare to the {\it COBE DMR}  results, we convert from multipole expansion
to the angular
correlation function with the {\it COBE} resolution \r\Smoot,
$$
   C(\theta) = {1\over4\pi}\sum_{l>2} \langle a_l^2 \rangle
      W_l^2 P_l(\cos\theta), \eqno(9)
$$
where
$$
   W_l^2 = \exp\biggl[-{l(l+1)\over17.8^2}\biggr]. \eqno(10)
$$
The cosmic variance of the correlation function is
$$
   \delta C(\theta)^2 = {1\over(4\pi)^2}\sum_{l>2} {2\over2l+1}
      \langle a_l^2\rangle^2 \bigl[W_l^2 P_l(\cos\theta)\bigr]^2. \eqno(11)
$$
We have found that it is sufficient to calculate multipoles up to $l = 40$.

\bigbreak
\centerline{\bf III. RESULTS}
\medskip

Fig.~1 shows the $53A+B \times 90A+B$ {\it COBE} angular correlation function
of Smoot
et al. \r\Smoot\ compared with the scale free spectrum,
i.e., $\vert \delta \vert^2 \propto \ k$.
We get a fit for a scale free spectrum (including both the cosmic
variance and the measurement uncertainties in the weighting) with a
minimum $\chi^2 = 48.0$.  For potentials of the type of eq.~(1) we also
get a minimum
$\chi^2 = 48.0$.  This occurs for $n \simls 0.1$.  These give spectra
almost identical to the scale-free one.  For potentials of the type of
eq.~(2), the extra parameter allows us to find a slightly lower $\chi^2
= 47.8$.
Since there are 70 measured
points, a two-parameter fit of the correct model
should have $\chi^2 \simls 68$.
Thus the $\chi^2$ per degree of freedom for the fit to the {\it COBE}
correlation function is exceedingly good and may indicate that there
is less scatter than one would
expect from a purely statistical distribution of cosmic variance and
experimental error.
We note that our minimum $\chi^2$ is slightly less than that quoted in
\r\Smoot~ for a pure power-law
spectrum with a Gaussian distribution of primordial fluctuations.  This
difference arises from restricting the current analysis to only
the $53A+B \times 90A+B$ cross correlation function which we had available.

For each $n$ or $(\alpha,\beta)$, we have done inflation calculations of
the fluctuation spectrum as a function of the potential amplitude, $\lambda$,
to find the best fit (minimum $\chi^2$) to the {\it COBE} correlation function.
Fig.~2 shows $\chi^2$ as a function of $n$ and contours of optimum
values for the amplitude $\lambda$ for chaotic inflation models
with effective potentials of the form
$$
   V(\phi) = {1\over n!}\lambda\phi^n~~~. \eqno(12)
$$
Note that larger values for $n$ give a worse fit to the {\it COBE} results.
A smaller $n$ means a smaller slope in the potential when the relevant scales
are generated, and thus a flatter (closer to scale free) spectrum.
Indeed, the correlation function for $n = 0.1$ (lowest value on Fig. 2) cannot
be distinguished
by eye from the completely scale-free case (Fig. 1).
Since the
scale-free spectrum fits the
{\it COBE} results so well, a larger deviation from the
scale-free spectrum obtained for the larger $n$ leads to a worse fit to the
COBE results.

In Fig.~3 we show the correlation
function for $n = 4$, which is the usually assumed self coupling \r\Linde~ in
chaotic inflation
models and still gives  a fairly good fit to the COBE results.
Fig.~4 shows an example of a poor fit for which $n = 1000$.  Here the
inflation-generated correlation function deviates significantly
from the {\it COBE} correlation function, particularly for angular scales
less than 50$^o$.

The confidence limits
for the parameters of these fits, $n$ and $\lambda$, are shown as
two-parameter contours of $\chi^2$  in Fig.~2.
For a two-parameter search, this just corresponds to the locus of points for
which
$\chi^2$ increases by 2.3 ($1 \sigma$ or 68\% C.L.)
or 6.2 ($2 \sigma$ or 95\% C.L.) as a function of $n$ and $\lambda$.  These
regions
include values of
$n \le 3.0$ ($1 \sigma$) and $n \le 22$ ($2 \sigma$).
The confidence limits on $n$ alone, allowing for {\it any} $\lambda$, are
however tighter \r\Numerical, corresponding to $\Delta\chi^2 = 1.0$ and
4.0.  Thus we obtain the upper limits $n \le 1.2$ ($1\sigma$ or 68\%
C.L.) and $n \le 7$ ($2\sigma$ or 95\% C.L.).
On the other hand, assuming a {\it fixed} $n$, $\lambda$ is determined
to better than 25\% ($1\sigma$), or 50\% ($2\sigma$) accuracy.

It is gratifying that a $\chi^2$ analysis does imply
optimum fits for relatively small coupling orders of the scalar field, and that
the
lower the order the better the fit.  This is consistent with what one expects
physically
since a self coupled scalar field in 4 dimensions is not renormalizable unless
 $n ^<_\sim 4$ \r\Coleman.

The most general renormalizable potential with just one
scalar field can be written,
$$
   V(\phi) = \lambda({\textstyle {1\over8}\beta\phi^2 + {1\over3}\alpha\phi^3
      +{1\over4}\phi^4)}~~. \eqno(13)
$$
Although more complicated forms including, for example, one-loop radiative
corrections are
 possible \r{\Coleman,\Bra,\Olive},
this form of the potential is sufficiently general for our purpose as it can
represent the leading terms in an expansion of a more complicated potential.
This term has been previously discussed by Hodges et al.~\r\Hodges~ in the
context  of
generating non-Zel'dovich spectra over scales of galactic clustering.
Here we apply it to the scales sampled by the {\it COBE} correlation function.

As in \r\Hodges, we can assume with no loss of generality that the initial
value for the scalar
field is greater than the global minimum.  This
is equivalent to a negative initial value and an $\alpha \rightarrow
-\alpha$ coordinate transformation.

Following Hodges et al., we can exclude immediately values for $\alpha$ and
$\beta$ such that,
$$
   \alpha>0, \beta<{\textstyle {8\over9}} \alpha^2~~~, \eqno(14)
$$
and
$$
   \alpha<0, {\textstyle 8\over9}\alpha^2 < \beta < \alpha^2~~~, \eqno(15)
$$
for which a false-vacuum secondary minimum of $V(\phi)$ occurs for positive
nonzero values of
the scalar field.
Such cases are excluded as, either the universe becomes
trapped in the false vacuum and can not exit inflation, or
(for a small false vacuum) they produce unacceptable
large-scale structure due to the wall energy associated with nucleated bubbles.

We present our results as a contour plot on the $(\alpha,\beta)$-plane
in Fig.~5.  As noted by Hodges et al. \r\Hodges, a significant deviation from a
scale-free spectrum occurs only as one approaches the line $\beta = \alpha^2$,
$\alpha < 0$
in Fig. 5.
This is the upper
thick line on the left of figure 5.  The thin lines on Fig. 5 identify contours
of optimum
values of $\lambda$ in the $\beta$-$\alpha$ plane as labeled.
 Also shown on Figure 5 is the shaded region of excluded parameter space
at the $1 \sigma$ level based upon the {\it COBE} data. The $2 \sigma$ 95.4\%
C. L.
excluded region is too small to distinguish from
the lines on Fig. 5.  We, therefore, show this region greatly expanded in
Fig.~6 reparameterized
\r\Hodges~ as $\beta' = (\beta- \alpha^2)/\alpha^6 $ vs. $\alpha$.

In the excluded region, there
is a flattening of $V(\phi)$ to the right of the global minimum.  This
extends the period of inflation near small values of $\phi$ where normally
little inflation
would occur.  This causes a deviation of the correlation function on the
largest
angular scales relative to the smallest scales.  This is  evidenced in Fig. 7
which shows
the predicted correlation function for the extreme case of  $\alpha = -0.9,~
\beta' = 0.00063$.

\bigbreak
\centerline{\bf IV. CONCLUSIONS}
\medskip
Although there remain large uncertainties from the cosmic variance
and from measurement errors of the angular correlation function for
fluctuations
in the microwave background,
we have shown that it is possible to place significant constraints on the
effective
potential for chaotic inflation based upon the {\it COBE DMR} data alone.
Based upon a $\chi^2$
analysis of the goodness of fit, and assuming a normal distribution of
measurement
errors and cosmic variance we conclude that only a chaotic inflation effective
potential
with relatively low orders of self coupling is consistent with the observed
correlation
function.  This in a small way adds credence to the chaotic inflation scenario
with
a single scalar field by the simple fact that, if the data had required a large
power of $\phi^n$,
 then the implied  potential would not have been renormalizable.  We have also
shown that
the physics of the underlying potential is constrained in that terms in the
effective potential which produce too much flattening  near the global minimum
are excluded by the data at the 1 and 2 $\sigma$ level.

Although most of this discussion has centered on the power-law for the
effective
potential, it is also worth noting that the strongest constraints are actually
upon the overall amplitude of the effective potential.  This is consistently
constrained
to be $^<_\sim 10^{-12}$, although larger values can occur for some polynomial
models (cf. Fig. 6).

Ultimately there is a limit as to how much the effective potentials can be
constrained due
to the unavoidable effects of cosmic variance. Indeed, it should
eventually be possible to identify the dispersion  in the correlation function
due to the cosmic variance alone. This would
 be a definitive confirmation of the inflationary scenario.
 Nevertheless, it is also clear that even stronger
constraints on the inflation-generating effective potential than those remarked
here
will be possible as the observed correlation function is better determined.

\bigbreak
\centerline{\bf V. ACKNOWLEDGEMENTS}
\medskip

We would like to acknowledge useful discussions with F. Graziani, N. J.
Snyderman,
and M. S. Turner.  We also are grateful to G. F. Smoot for providing us with
the {\it COBE DMR} $53A+B \times 90A+B$ angular correlation function.
Computations were carried out in part at the Centre for Scientific Computing,
Finland.  HK-S thanks the Academy of Finland for financial support.
Work performed in part under the auspices of the U.~S.~Department of Energy
by the Lawrence Livermore National Laboratory under contract
W-7405-ENG-48.

\bigbreak
\centerline{\bf References}
\medskip

\def\i#1{\par\hang\indent\llap{[#1]\quad}\ignorespaces}
{\frenchspacing
\i\Smoot   G. F. Smoot et al., Ap. J. {\bf 396}, L1 (1992);
           E. L. Wright et al., Ap. J. {\bf 396}, L13 (1992).
\i\Davis   R. L. Davis, H. M. Hodges, G. F. Smoot, P. J. Steinhardt, and M. S.
Turner,
Phys. Rev. Lett., {\bf 69}, 1856 (1992).
\i\Bra     R. Brandenberger, R. Kahn, and W. H. Press, Phys. Rev. D {\bf 28},
           1809 (1983); R. Brandenberger and R. Kahn, Phys. Rev. D {\bf 29},
           2172 (1984); R. H. Brandenberger, Rev. Mod. Phys. {\bf 57}, 1
(1985);
           A. H. Guth and S.-Y. Pi, Phys. Rev. Lett. {\bf 49}, 1110 (1982);
           J. M. Bardeen, P. J. Steinhardt, and M. S. Turner, Phys. Rev.
           D {\bf 28}, 679 (1983).

\i\Olive  K. A. Olive, Phys. Rep., {\bf 190}, 307 (1990).
\i\Copeland  E. J. Copeland, E. W. Kolb, A. R. Liddle, and J. E. Lindsey, Phys.
Rev {\bf D}
submitted (1993).
\i\Adams   F. C. Adams, J. R. Bond, K. Freese, J. A. Frieman, and A. V. Olinto,
Phys. Rev., {\bf D47}, 426 (1993).
\i\AbbWis  L. F. Abbot and M. B. Wise, Nucl. Phys., {\bf B244}, 541 (1984).
\i\Linde A. Linde, Phys. Lett., {\bf B129}, 177 (1983).
\i\Coleman  S. Coleman and E. Weinberg, Phys. Rev., {\bf D7}, 1888 (1973).
\i\Numerical  W. Press, B. P. Flannery, S. A. Teukolski, and W. T. Vetterling,
{\it Numerical Recipes}, (Cambridge: Cambridge University Press) (1986).
\i\Bevington P. R. Bevington, {\it Data Reduction and Analysis for the Physical
Sciences},
(New York: McGraw Hil) 1969.
\i\Hodges  H. M. Hodges, G. R. Blumenthal, L. A. Kofman, and J. R. Primack,
Nucl. Phys., {\bf B335}, 197 (1990).
}

\vfil\eject


\bigbreak
\centerline{\bf Figure Captions}
\medskip

FIG.~1.  The {\it COBE DMR} $53A+B \times 90A+B$ angular correlation function
(points)
compared with an inflation-generated scale-free spectrum (central solid line).
The upper
and lower solid lines illustrate the effects of cosmic variance.
\medskip

FIG.~2.  Computed $\chi^2$ (thick line) and optimum $\lambda$ (dashed line)
as a function of $n$ for a chaotic
inflation potential with $V(\phi) = \lambda \phi^n$. The thin lines enclose
the 68\% and 95\% confidence regions in the  $(n,\lambda)$ parameter plane.
The scales for $\chi^2$ and $\lambda$ are
on the left and right of this figure respectively.
\medskip

FIG.~3.  The {\it COBE DMR} $53A+B \times 90A+B$ angular correlation function
(points)
compared with a chaotic inflation  $V(\phi) = \lambda \phi^4$ spectrum (central
solid line).
The upper and lower solid lines illustrate the effects of cosmic variance.
\medskip

FIG.~4.  The {\it COBE DMR} $53A+B \times 90A+B$ angular correlation function
(points)
compared with a chaotic inflation  $V(\phi) = \lambda \phi^n$ spectrum (central
solid line),
where $n = 1000$.
The upper and lower solid lines illustrate the effects of cosmic variance.
\medskip

FIG.~5.  Excluded regions of the $\alpha$ vs. $\beta$ plane for a polynomial
chaotic inflation potential.  The lightly shaded region shows the excluded
parameter
space at the $1 \sigma$ level based upon fits to the {\it COBE DMR} results. As
in
\r\Hodges~ we also exclude the darker shaded regions with $\beta < 8
\alpha^2/9$,
$\alpha > 0$ (lower right),
and $8 \alpha^2/9 < \beta < \alpha^2$
(upper left), for which
there is a secondary minimum in $V(\phi)$ for positive $\phi$.
The thin lines show
contours of optimum $\lambda$ for each ($\alpha, \beta$) as labeled.

\medskip

FIG.~6.
Region of excluded parameter
space (shaded region) at the $2 \sigma$ level based upon fits to the {\it COBE
DMR} results
using a polynomial chaotic inflation potential.
Parameters are plotted in the expanded $\alpha$ vs. $\beta' =
(\beta-\alpha^2)/\alpha^6$
plane.  The thin lines show
contours of optimum $\lambda$ for each ($\alpha, \beta$) as labeled.

\medskip

FIG.~7.  The {\it COBE DMR} $53A+B \times 90A+B$ angular correlation function
(points)
compared with a chaotic inflation spectrum (central solid line) generated with
a polynomial effective potential with parameters $\alpha$ and $\beta'$ as
indicated.
\medskip

\bye